%Paper: gr-qc/9207004
%From: CANGEMI@lnstmp.mit.edu (Daniel Cangemi)
%Date: Thu, 23 Jul 1992 10:08:55 -0400 (EDT)

%
% This is a Plain TeX file
%
\def\IR{\relax{\rm I\kern-.18em R}}
\def\ll{\left\langle}
\def\rr{\right\rangle}
\def\diag{\,{\rm diag}\,}
\magnification=1200
\hoffset=-.1in
\voffset=-.2in

\vsize=7.5in
\hsize=5.6in
\tolerance 10000

\baselineskip 12pt plus 1pt minus 1pt
\pageno=0
\centerline{\bf ONE FORMULATION FOR BOTH LINEAL GRAVITIES}
\centerline{{\bf THROUGH A DIMENSIONAL REDUCTION}\footnote{*}{This
work is supported in part by funds
provided by the U. S. Department of Energy (D.O.E.) under contract
\#DE-AC02-76ER03069, and the Swiss National Science Foundation.}}
\vskip 24pt
\centerline{D.~Cangemi}
\vskip 12pt
\centerline{\it Center for Theoretical Physics}
\centerline{\it Laboratory for Nuclear Science}
\centerline{\it and Department of Physics}
\centerline{\it Massachusetts Institute of Technology}
\centerline{\it Cambridge, Massachusetts\ \ 02139\ \ \ U.S.A.}
\vskip 1.5in
\centerline{Submitted to: {\it Physics Letters B\/}}
\vfill
\centerline{ Typeset in $\TeX$ by Roger L. Gilson}
\vskip -12pt
\noindent CTP\#2124\hfill July 1992
\eject
\baselineskip 24pt plus 2pt minus 2pt
\centerline{\bf ABSTRACT}
\medskip
The two lineal gravities --- based on the de~Sitter group or a central
extension of the Poincar\'e group in $1+1$ dimensions --- are shown to derive
classically from a unique topological gauge theory.  This one is obtained
after a dimensional reduction of a Chern--Simons model, which describes pure
gravity in $2+1$ dimensions, the gauge symmetry being given by an extension of
$ISO(2,1)$.
\vfill
\eject
\noindent{\bf INTRODUCTION}
\medskip
There is increasing interest these last years in studying gravity in
low-dimensional space-times.  In $2+1$ dimensions the Einstein equations
undergo a drastic simplification which allows a deeper understanding of global
gravitational effects.$^{1}$  Moreover, in the absence of matter, they
were shown$^2$ to be the equations
of motion of a topological Chern--Simons theory.
 According to the sign of the cosmological constant $\Lambda$, the underlying
gauge group is usually taken to be $ISO(2,1)$ for $\Lambda=0$ (the Poincar\'e
group), $SO(3,1)$ for $\Lambda>0$ (the de~Sitter group) or $SO(2,2)$ for
$\Lambda<0$ (the anti-de~Sitter group).

In $1+1$ dimensions, in spite of the vanishing of the Einstein tensor, two
alternatives were proposed as lineal gravities.
Both models were shown to be classically --- and perhaps also quantically$^3$
--- equivalent to a topological gauge theory.  Their actions have
the general form:
$$L = \int_{\Sigma} \ll H,F\rr\ \ ,\eqno(1)$$
where $F$ is a (curvature) two-form, $H$ is a Lagrange multiplier scalar
function --- both take value in a Lie algebra and transform according to the
adjoint representation
--- and $\ll~~,~~\rr$ defines an
invariant non-degenerate bilinear form on this Lie algebra.$^4$

In the first case,$^5$ the gauge symmetry is given by the (anti-) de~Sitter
$SO(2,1)$ group:
$$[P_a, J] = \epsilon_a{}^b P_b\ \ ,\qquad [P_a, P_b] = - {\Lambda\over 2}
\epsilon_{ab} J \ \ .\eqno(2)$$
Using the inner product coded in the Casimir $P_a P^a + {\Lambda\over 2} J^2$
and writing $A = e^a P_a + \omega J$, $H = \eta^a P_a + {\Lambda\over
2}\eta_2J$, the Lagrange density in Eq.~(1) is:
$${\cal L} = \eta_a \left( de^a + \epsilon^a{}_b \omega e^b\right) +
\eta_2 \left( d\omega - {1\over 4}\Lambda \epsilon_{ab} e^a e^b\right)\ \
.\eqno(3)$$
[The indices $(a,b,\ldots)$ take the value 0,1 and are raised or lowered with
the flat-space metric $\eta_{ab} = \diag (1,-1)$.  $\epsilon^{ab}$ is the
anti-symmetric tensor with $\epsilon^{01}=1$.]  The components of the gauge
field
$e^a,\omega$ are interpreted
as the {\it Zweibein\/} and the spin-connection one-forms.
The Lagrange multiplier functions $\eta_a$, $\eta_2$
enforce the scalar curvature to be equated
to the constant $\Lambda$.  This is called the {\it de~Sitter\/} model.

On the other hand,
the gauge symmetry of the second proposal
was recently$^{6}$ identified as a central extension of the
Poincar\'e algebra:
$$[P_a,J] = \epsilon_a{}^b P_b\ \ ,\qquad [P_a,P_b] = \epsilon_{ab}I\ \
,\qquad [I,J] = 0 = [I,P_a]\ \ .\eqno(4)$$
Therefore we shall call this model the {\it extended
Poincar\'e\/} model.  With the inner product given by the Casimir $P_a P^a -
JI - IJ$ and the decomposition $A = e^a P_a + \omega J + aI$, $H = \eta^a P_a
- \eta_3 J - \eta_2 I$, the Lagrangian density in Eq.~(1) becomes:
$${\cal L} = \eta_a \left( de^a + \epsilon^a{}_b \omega e^b\right) +
\eta_2 d\omega + \eta_3 \left( da + {1\over 2} \epsilon_{ab} e^a e^b\right)\ \
.\eqno(5)$$
Remark that now one of the equations of motion set the scalar curvature to
zero and that $\eta_3$ plays the role of a cosmological constant.$^7$ A
non-conventional contraction from the de~Sitter algebra to the extended
Poincar\'e one relates both models.  Namely, adding a $U(1)$ generator $I$ in
the algebra (2), replacing $J$ by $J-2I/\Lambda$ and taking the limit
$\Lambda\to 0$ lead to the algebra (4).

If it was already remarked that the de~Sitter model can be viewed as a
dimensional reduction of a Chern--Simons
model with $2+1$-de~Sitter gauge group,
it was not clear whether the extended Poincar\'e theory follows from a
similar reduction.  We show in this paper that this can indeed be
achieved provided we start in $2+1$ dimensions from an extension of the
Poincar\'e $ISO(2,1)$ group.  We first
propose a topological gauge theory based on a new symmetry as pure gravity in
$2+1$ dimensions.  The dimensionally reduced equations of motion admit
then not only the extended Poincar\'e
classical solutions but also the de~Sitter ones.
We thus get a unified picture of the two different lineal gravities.

In Section~I we construct the minimal extension of the $2+1$-Poincar\'e
algebra which contains the extended $1+1$-Poincar\'e one.  In Section~II we
perform a dimensional reduction and we
derive the general equations of motion we are left
with. They admit Einstein-type solutions with an
arbitrary cosmological constant.  In Section~III we show that the de~Sitter
and extended Poincar\'e
classical solutions are among them.  Finally we give some
comments and conclusions in Section~IV.
\goodbreak
\bigskip
\hangindent=20pt \hangafter=1
\noindent{\bf I.\quad THE MINIMAL EXTENSION OF ISO(2,1) AND 2~+~1-GRAVITY}
\medskip
\nobreak
It is always interesting to look at a theory as the dimensional reduction of
another one.  The extra space-time dimensions carry additional information
describing, for example, electromagnetism in the Kaluza-Klein model$^8$ or a
Higgs field in the Manton--Meyer model.$^9$
In our case we look for gravity theories in
$2+1$ dimensions which reduce in $1+1$ dimensions to the two lineal gravities
we have just described.  The dimensional reduction that will be explicitly
shown in the next section consists of imposing translational invariance along
a spatial direction.  Contrary to the cited cases, this reduction does not
change the gauge symmetry and hence the gauge group is not modified.

The last remark has lead us to search for a group containing the extended
Poincar\'e one (4) and which still gives a description of gravity in $2+1$
dimensions.  This can be achieved by an extension of $ISO(2,1)$ similar to the
one of $ISO(1,1)$.
We notice that a central extension of $ISO(2,1)$ does not exist.
Let us consider the (anti) de~Sitter algebra in $2+1$
dimensions:
$$\left[ \bar{J}_A, \bar{J}_B\right] = \epsilon_{AB}{}^C \bar{J}_C\ \ ,\qquad
\left[ \bar{J}_A, P_B\right] = \epsilon_{AB}{}^C P_C\ \ ,\qquad
\left[ P_A, P_B\right] =- \Lambda\epsilon_{AB}{}^C \bar{J}_C \eqno(6)$$
[the indices $(A,B,C,\ldots)$ take the value $0,1,2$ and are moved with
the metric $\eta_{AB} = \diag (1,-1,-1)$.  $\epsilon^{ABC}$ is the totally
antisymmetric tensor with $\epsilon^{012}=1$].  We extend it trivially by
$SU(2)$:
$$\left[ S_A, S_B\right] = \epsilon_{AB}{}^C S_C\ \ ,\qquad \left[ S_A,
\bar{J}_B\right] = 0\ \ ,\qquad \left[ S_A, P_B\right] = 0 \eqno(7)$$
and we perform the contraction, $J_A = \bar{J}_A + S_A$, $I_A=-\Lambda S_A$,
$\Lambda\to 0$:
$$\eqalign{
\left[ J_A, J_B\right] &= \epsilon_{AB}{}^C J_C\ \ ,\cr
\left[ J_A, P_B\right] &= \epsilon_{AB}{}^C P_C\ \ ,\qquad \left[ P_A,
P_B\right] = \epsilon_{AB}{}^C I_C\ \ ,\cr
\left[ J_A, I_B\right] &= \epsilon_{AB}{}^C I_C\ \ ,\qquad \left[ P_A,
I_B\right] = 0\ \ ,\qquad \left[ I_A, I_B\right] = 0\ \ .\cr}\eqno(8)$$
This algebra is an extension of the Poincar\'e algebra by an Abelian ideal (of
dimension three).

To write the Chern--Simons action with the gauge symmetry (8) we need
an invariant non-degenerate bilinear form.  The most general one
is easily shown to be parametrized by two real constants $c_1,c_2$.  In the
basis $\{T_M\}^9_{M=1}=\{P_A, J_A, I_A\}^2_{A=0}$ we get:
$$\ll T_M, T_N\rr \equiv h_{MN} = \left( \matrix{
\eta_{AB} & c_2\eta_{AB} & 0 \cr
c_2 \eta_{AB} & c_1 \eta_{AB} & \eta_{AB} \cr
0 & \eta_{AB} & 0 \cr}\right) \eqno(9)$$
which is associated with the Casimir:
$$C = P_A P^A - c_2 \left( P_A I^A + I_A P^A\right) + \left( J_A I^A + I_A
J^A\right) + \left( c^2_2 - c_1\right) I_A I^A \ \ .\eqno(10)$$
With $\tilde A = e^A P_A + \omega^A J_A + a^A I_A$ a Lie algebra valued
one-form, our action is written:
$$\eqalign{\tilde L &=  \int_{\IR^3} \ll \tilde A,
d\tilde A + {2\over 3}\tilde A^2\rr = \int d^3x\, \tilde{\cal L}
\cr\noalign{\vskip 0.3cm}
\tilde{\cal L} &= 2c_2 e_A \left( d\omega^A + {1\over 2}
\epsilon^A{}_{BC} \omega^B \omega^C\right) + e_A \left( de^A +
\epsilon^A{}_{BC} \omega^B e^C\right) \cr
&+ 2 a_A \left( d\omega^A + {1\over 2} \epsilon^A{}_{BC} \omega^B \omega^C
\right) + c_1 \left( \omega_A d\omega^A + {1\over 3} \epsilon_{ABC} \omega^A
\omega^B \omega^C \right)\ \ .\cr}\eqno(11)$$
(We have dropped in $\tilde{\cal L}$ some exact differentials contributing by
surface terms only.)
If $e^A, \omega^A$ are interpreted as a {\it Dreibein\/}
and a spin-connection, the
first term is just the scalar curvature and the second one the scalar
torsion.$^{10}$

The relation with pure gravity is still clearer if we present the equations
of motion derived from the action (11).  For arbitrary $c_1,c_2$,
the equations of motion
are always given by the zero curvature condition, $d\tilde A + \tilde A^2 =
0$; in components we get:
$$\eqalign{
&de^A + \epsilon^A{}_{BC} \omega^B e^C = 0 \ \ ,\cr
&d\omega^A + {1\over 2} \epsilon^A{}_{BC} \omega^B \omega^C = 0\ \ ,\cr
&da^A + \epsilon^A{}_{BC} \omega^B a^C + {1\over 2} \epsilon^A{}_{BC} e^B e^C
= 0\ \ .\cr}\eqno(12)$$
We recognize in the two first equations the torsion free and the Einstein
equations of pure gravity without cosmological constant.  The role of the
field $a^A$ is still to be elucidated.  If it appears only as an auxiliary
field at the classical level, its presence in the Lagrangian (11) could have a
decisive contribution in the quantization procedure and induce different
results than the usual model.$^{2}$  Thus Eq.~(11) is an alternative to
pure gravity
(without cosmological constant) based on an extension of $ISO(2,1)$.
\goodbreak
\bigskip
\noindent{\bf II.\quad THE DIMENSIONALLY REDUCED MODEL}
\medskip
\nobreak
We now descend to $1+1$ dimensions by imposing a translational invariance
along the second spatial direction.  It is useful to write the one-form
$\tilde A = A + A_2 dx^2$ and to introduce the curvature two-form $F=dA +
A^2$.  Apart from a surface term and an (infinite) constant generated by the
integration along $x^2$, the action (11) is reduced to:
$$L = \int_{\IR^2} \ll A_2, F\rr\ \ .\eqno(13)$$
If we identify $A_2$ with the scalar function $H$, we recognize the topological
action (1) in $1+1$ dimensions.
Using the inner product (9) in Eq.~(13) we can rearrange the
terms with the help of the convenient definitions:
$$A_{2,M} \equiv
 h_{MN} A^N_2 \equiv \left( \left( \eta^{(1)}_a, \eta^{(1)}\right), \left(
\eta^{(2)}_a, \eta^{(2)}\right), \left( \eta^{(3)}_a, \eta^{(3)}\right)\right)
\eqno(14)$$
[we recall that $(a,b,\ldots)$ takes the value 0,1 only and are moved with the
metric $\diag(1,-1)$].  Due to the invertibility of the inner product,
these fields
are independent.  If the reduced connection $A$ ($\tilde A = A + A_2 dx^2$) is
decomposed according to:
$$A = e^a P_a + eP_2 + \omega^a J_a + \omega J_2 + a^a I_a + aI_2 \eqno(15)$$
the Lagrangian density in (13) becomes:
$$\eqalign{{\cal L}&= \eta^{(1)}_a \left( de^a + \epsilon^a{}_b\omega e^b +
\epsilon^a{}_b e\omega^b\right) + \eta^{(2)}_a \left( d\omega^a +
\epsilon^a{}_b \omega\omega^b\right) \cr
&+ \eta^{(3)}_a \left( da^a + \epsilon^a{}_b \omega a^b + \epsilon^a{}_b ee^b
+ \epsilon^a{}_b a\omega^b\right) \cr
&+ \eta^{(1)} \left( de + \epsilon_{ab} \omega^a e^b\right) + \eta^{(2)}
\left( d\omega + {1\over 2} \epsilon_{ab} \omega^a \omega^b\right) \cr
&+ \eta^{(3)} \left( da + {1\over 2} \epsilon_{ab} e^a e^b + \epsilon_{ab}
\omega^a a^b\right)\ \ ,\cr}\eqno(16)$$
and will be called the {\it reduced\/} Lagrangian.

Interpreting $e^a$ as a {\it Zweibein\/} and $\omega$ as a
spin-connection, we recognize the torsion tensor $de^a + \epsilon^a{}_b \omega
e^b$ multiplying $\eta^{(1)}_a$ and the scalar curvature $d\omega$ multiplying
$\eta^{(2)}$.  For a theory of gravitation we want the {\it Zweibein\/}
and the spin connection to be related --- at least classically --- by a
torsion free condition.  This is obtained by varying (16) with respect to
$\eta^{(1)}_a$ if and only if either $e$ or $\omega^a$ is identically zero.

Let us look first if such a choice is consistent with the other equations of
motion.  The ones involved are obtained by variations with respect to
$\eta^{(1)}$ and $\eta^{(2)}_a$:
$$\eqalignno{de + \epsilon_{ab} \omega^a e^b &= 0 &(17\hbox{a})\cr
d\omega^a + \epsilon^a{}_b \omega \omega^b &= 0\ \ .&(17\hbox{b}) \cr}$$
Try first $e\equiv 0$.  Then (17a) implies for $\omega^a$:
$$\omega^a = \sqrt{{\Lambda\over 2}} \epsilon^a{}_b e^b \eqno(18)$$
with an arbitrary positive constant $\Lambda$ (possibly set to zero).  Then
(17b) gives just a multiple of the torsion free condition and $e\equiv 0$ is a
consistent choice

By a variation with respect to $\eta^{(1)}_a$ and $\eta^{(2)}$ and using the
previous solution for $e$ and $\omega^a$ we get:
$$de^a + \epsilon^a{}_{b} \omega e^b = 0\ \ ,\qquad d\omega - {1\over
4}\Lambda \epsilon_{ab} e^a e^b = 0 \eqno(19)$$
which are the basic equations of the de~Sitter model if $\Lambda\not=0$ and of
the extended Poincar\'e model if $\Lambda \equiv 0$.
The choice $\omega^a=0$ is also
consistent and implies $e$ to be constant [cf.~Eq.~(17a)].  Equation (19) holds
with $\Lambda=0$.

We conclude this section by emphasizing the result (19).  Einstein-type
gravities with arbitrary cosmological constant are among the solutions of the
reduced equations of motion [coming from Lagrangian (16)].  They are
characterized by a vanishing torsion tensor.  Another intriging
point is the
gauge structure of Lagrangian (16).  After the reduction, the symmetry is still
given by the extended $2+1$-Poincar\'e  algebra.  Even if this seems not to be
a ``natural'' algebra in $1+1$ dimensions, the extended Poincar\'e
{\it and\/} the de~Sitter gauge structures
 can be recovered on-shell as we shall show in the next
section.
\goodbreak
\bigskip
\hangindent=29pt \hangafter=1
\noindent{\bf III.\quad THE EXTENDED POINCAR\'E
AND THE DE~SITTER MODELS IN LIGHT OF THE REDUCED MODEL}
\medskip
\nobreak
We present here how the classical solutions of the extended Poincar\'e and the
de~Sitter actions lie among the classical solutions of the
reduced action. More precisely, we show how special {\it
Ans\"atze\/} reduce the equations coming from (16) to those coming from (5) or
(3).

The extended $1+1$-Poincar\'e algebra (4) has a natural
embedding in the extended $2+1$-Poincar\'e algebra (8).
One checks that the generators:
$$P'_a = P_A\ \ ,\qquad J' = J_2\ \ ,\qquad I' = I_2\eqno(20)$$
span in the algebra (8) a subalgebra which is exactly the
one given by Eq.~(4).  On the other hand, the linear combinations
$$\eqalign{P'_a &= P_a - \sqrt{{\Lambda\over 2}} \epsilon_a{}^b J_b - {1\over
\sqrt{2\Lambda}} \epsilon_a{}^b I_b\ \ ,\cr
J' &= J_2 \cr}\eqno(21)$$
generate the de~Sitter algebra (2) in $1+1$ dimensions.  In other words we
notice that the extended $2+1$-Poincar\'e algebra contains both the extended
Poincar\'e and de~Sitter algebras.

Moreover, the inner product (9) induces in these subalgebras
the general invariant inner
products of the de~Sitter and the extended Poincar\'e algebras.
This means that restricting
 the gauge field in Eq.~(16) to one of the algebras (20)
or (21) we get the usual Lagrangian densities (3) or (5).  Their classical
solutions are thus among the ones of the reduced model.  More explicitly, the
{\it Ansatz\/}:
$$\eqalign{A &= e^a \left( P_a - \sqrt{{\Lambda\over 2}} \epsilon_a{}^b J_b -
{1\over \sqrt{2\Lambda}}\epsilon_a{}^b I_b\right) + \omega J_2\ \ ,\cr
H &= \eta^a \left( P_a - \sqrt{{\Lambda\over 2}} \epsilon_a{}^b - {1\over
\sqrt{2\Lambda}} \epsilon_a{}^b I_b\right) + {\Lambda\over 2} \eta_2 J_2\ \ ,
\cr}\eqno(22)$$
allows us to
solve the equations of motion of the reduced model and the solutions coincide
with the de~Sitter ones if $e^a$, $\omega$, $\eta^a$, $\eta_2$ correspond to
the fields in
(3).  The anti-de~Sitter model (with $\Lambda$ negative) can be treated in a
similar way, but we do not present the details here.  The same remark applies
to the extended Poincar\'e model with the {\it Ansatz\/}:
$$\eqalign{A &= e^a P_A + \omega J_2 + aI_2\ \ ,\cr
H &= \eta^a P_A + \eta_3 J_2 + \eta_2 I_2\ \ .\cr}\eqno(23)$$
Here $e^a$, $\omega$, $a$, $\eta^a$, $\eta_2$, $\eta_3$ are identified with
the corresponding fields in (5).
\goodbreak
\bigskip
\noindent{\bf IV.\quad COMMENTS AND CONCLUSIONS}
\medskip
We have shown that the equations of motion of the
reduced model (16) possesses both the de~Sitter
($\Lambda\not=0$) and the extended Poincar\'e ($\Lambda=0$) solutions.
If we consider a trivial bundle on
$\Sigma=\IR^2$,
the space of classical solutions of the topological model (1) is given by
all the flat connections.  This space is contractible and if we divide it by
all the gauge transformations it reduces to a point.  In order to interpret
(16) as a model of gravity, we identify in the decomposition (15) $e^a$ as a
{\it Zweibein\/} and
construct a metric in the usual way $g_{\mu\nu} = \eta_{ab} e^a_\mu
e^b_\nu$.  But we have to impose that $-{1\over 2} \epsilon_{ab} e^a e^b$
never vanishes since it is proportional to $\sqrt{-\det\,g}$.  This means
that not all configurations are geometrical solutions of the equations of
motion.   The configurations with non-vanishing $-{1\over 2}\epsilon_{ab} e^a
e^b$ form a set which no longer has a
trivial topology.  For example, in our reduced model, the de~Sitter and the
extended Poincar\'e solutions are clearly disconnected.$^{11}$

In this paper we have obtained the two lineal gravities as a reduction of a
$2+1$-dimensional topological theory.  This one is in fact a model of
$2+1$-gravity whose symmetry is an extension of the Poincar\'e algebra.  The
reduction to $1+1$-dimensions gives an action whose classical solutions can be
the de~Sitter or the extended Poincar\'e ones.
The quantization of this model will deserve further study.
\goodbreak
\bigskip
\centerline{\bf ACKNOWLEDGEMENTS}
\medskip
I thank Roman Jackiw for helpful comments and Vesa Ruuska for many discussions
on the mathematical aspects of this work.
\vfill
\eject
\centerline{\bf REFERENCES}
\medskip
\item{1.}S. Deser, R. Jackiw and G. 't~Hooft, {\it Ann. Phys.\/} (NY) {\bf
152}, 220 (1984); for a recent example, see S.~Deser and R.~Jackiw, ``Time
Travel?,'' , Brandeis University preprint
BRX~TH~334 (June 1992) and MIT preprint\#2101 (June 1992).
\medskip
\item{2.}A. Achucarro and P. Townsend, {\it Phys. Lett.\/} {\bf 180B}, 89
(1986); E. Witten, {\it Nucl. Phys.\/} {\bf B311}, 46 (1988/89).
\medskip
\item{3.}H. Terao, ``Quantum Analysis of Jackiw and Teitelboim's Model for
$1+1$-Dimensional Gravity and Topological Gauge Theory,'' Kanazawa preprint
DPKU-9207 (April 1992).
\medskip
\item{4.}The
topological action (1) with (semi-simple) Lie algebras was also considered by
K. Aoki and E. D'Hoker in ``$W$ Gravity and Generalized Lax Equations for
(Super) Toda Theory,'' preprint UCLA/92/TEP/12.
\medskip
\item{5.}T. Fukayama and K. Kamimura, {\it Phys. Lett.\/} {\bf 160B}, 259
(1985); K. Isler and C. A. Trugenberger, {\it Phys. Rev. Lett.\/} {\bf 63},
834 (1989); A. H. Chamseddine and D. Wyler, {\it Phys. Lett.\/} {\bf 228B}, 75
(1989).
\medskip
\item{6.}D. Cangemi and R. Jackiw, ``Gauge Invariant Formulations of Lineal
Gravities,'' {\it Phys. Rev. Lett.\/} {\bf 69}, 233 (1992).
\medskip
\item{7.}R. Jackiw, in {\it Proceedings of the XIX International Colloquium on
Group Theoretical Methods in Physics\/}, Salamanca, Spain, June 29--July 4,
1992, to be published; (MIT preprint CTP\#2125 (July 1992).)
\medskip
\item{8.}T. Kaluza, {\it Sitz, Preuss. Akad.\/} {\bf 966} (1921); O. Klein,
{\it Z. Phys.\/} {\bf 37} (1926).
\medskip
\item{9.}N. Manton, {\it Nucl. Phys.\/} {\bf B158}, 141 (1979); M. Meyer
(unpublished) [University of California at Irvine preprint].
\medskip
\item{10.}Note that in the limit $c_2\to\infty$, $c^{-1}_2\tilde{\cal L}$ is
the
action of pure gravity based on $ISO(2,1)$ and given in Ref.~[2].
\medskip
\item{11.}A review of the classical
solutions of the de~Sitter and the extended
Poincar\'e models can be found in R. Jackiw, ``Gauge Theories for Gravity on a
Line,'' MIT preprint CTP\#2105 (June 1992).
\par
\vfill
\end